\documentclass[conference]{IEEEtran}

\usepackage[T1]{fontenc}

\usepackage[cmintegrals]{newtxmath}
\usepackage{epsfig}
\usepackage{afterpage}
\usepackage{graphicx}
\usepackage{amsmath}
\usepackage{kotex}

\usepackage{cases}
\usepackage{subcaption}

\usepackage{amsfonts}
\usepackage{cite}
\usepackage{times}
\usepackage{color}
\usepackage{cite}
\usepackage{psfrag}
\usepackage{dblfloatfix}
\usepackage{multirow}
\usepackage{cuted,flushend}
\usepackage{mathtools}
\usepackage{float}
\usepackage{array}

\begin{document}

 \IEEEoverridecommandlockouts 
 

\bstctlcite{IEEEexample:BSTcontrol}

\title{Graph Neural Network Based Beamforming and RIS Reflection Design in A Multi-RIS Assisted Wireless Network 
\vspace{-3mm}\thanks{This work was supported in part by the National Science Foundation under CNS grant 1908552 and ECCS Grant 1808912.}\vspace{-3mm}
}

\author{\IEEEauthorblockN{Byungju Lim\IEEEauthorrefmark{1} and Mai Vu\IEEEauthorrefmark{2}}
\IEEEauthorblockA{\IEEEauthorrefmark{1}Department of Electronic Engineering,
Pukyong National University, Busan, South Korea}
\IEEEauthorblockA{\IEEEauthorrefmark{2}Department of Electrical and Computer Engineering, Tufts University, Medford, USA}
\IEEEauthorblockA{Email: limbj@pknu.ac.kr, mai.vu@tufts.edu}
}

\maketitle
\vspace{-3mm}
\begin{abstract}
We propose a graph neural network
(GNN) architecture to optimize base station (BS) beamforming
and reconfigurable intelligent surface (RIS) phase shifts in a
multi-RIS assisted wireless network. We create a bipartite graph
model to represent a network with multi-RIS, then construct
the GNN architecture by exploiting channel information as
node and edge features. We employ a message passing
mechanism to enable information exchange between
RIS nodes and user nodes and facilitate the inference of
interference. Each node also maintains a representation vector
which can be mapped to the BS beamforming or RIS phase shifts
output. Message generation and update of the representation
vector at each node are performed using two unsupervised
neural networks, which are trained off-line and then used on
all nodes of the same type. Simulation results demonstrate
that the proposed GNN architecture provides strong scalability
with network size, generalizes to different settings, and
significantly outperforms conventional algorithms.

\end{abstract}



\vspace{-2mm}
\section{Introduction}\label{sec:Intro}

Reconfigurable intelligent surface (RIS) technology promises to enhance wireless network performance by supporting each wireless communication link with an additional path. An RIS reflects the intended signal to the desired receiver but also reflects that signal to unexpected receivers, hence introducing new interference to these receivers. Furthermore, the presence of RISs in a network changes the wireless propagation and thus affects the optimal BS beamforming. Thus, it is challenging to design RIS phase shifts and BS beamforming simultaneously in a multi-RIS assisted network.
Enhancing the performance of desired receiver while suppressing interference to other receivers should be considered.

Recent research efforts have focused on the optimization of BS beamforming and RIS phase shifts using conventional convex optimization techniques \cite{ris_basic,ris_basic2,ris_basic3}.
Deep neural network (DNN) or deep reinforcement learning (DRL) based approach has also been considered \cite{JSAC_RIS_DRL,TCCN_RIS_DL,IoT_RIS_DL}, where
the DNN approach can provide a good performance with low complexity while DRL exhibits high adaptability to the changes in a wireless environment.
However, DRL can have slow convergence, whereas the DNN learning approach has poor scalability and generalization since the input size of a DNN is on the order of the number of nodes in a wireless network.
If several nodes are removed or added in a network, the conventional deep learning network needs to be re-designed and re-trained under a different network setting.
To achieve a better scalability with network size, the graph neural network (GNN) approach has been applied \cite{TWC_bipartite,RISGNNJSAC2021}.
The principal of GNN is to apply the same neural network at each node after modeling a wireless network into a graph, and train the neural network with the same set of weight parameters.

GNN has been used to perform the joint optimization of beamforming and RIS phase shift in a singe-RIS wireless network using implicit channel information \cite{RISGNNJSAC2021}.
This reference shows that GNN based optimization provides good scalability and generalization when the number of users in the network changes.
However, the GNN structure considered in \cite{RISGNNJSAC2021} does not have any feature and also cannot directly exploit the relationship between nodes.
Furthermore, it can be only applied to networks with a single RIS.

In this paper, we propose a novel GNN structure to optimize BS beamforming and RIS phase shifts in a wireless network aided by multiple RISs.
We first represent a general graph model for a wireless network with multi-RIS.
We then use this graph model to design a GNN with node and edge features as channel information, and introduce a message passing mechanism for sharing information between nodes.
The message passing exploits the relationship between nodes using explicit channel information and facilitates the learning of interference at each user node. 
Numerical results show that the proposed GNN structure provides strong scalability and generalization over a varying change of number of RISs.
Further, it achieves superior network sum rate performance over conventional methods.


\vspace{-2mm}
\section{System and Signal Models}\label{sec:sys}

\subsection{System and Signal models}
We consider a multi-RIS assisted single-cell network where a base station (BS) serves $K$ single-antenna users with the aid of $J$ RISs.
Assume that the BS is equipped with $N$ antennas and each RIS is equipped with $M$ passive reflecting elements.

Let $\mathbf{h}_{k}\in \mathbb{C}^{N\times 1}$ denote the channel between the BS and user $k$, $\mathbf{f}_{j,k}\in \mathbb{C}^{M\times 1}$ denote the channel from RIS $j$ to user $k$, and $\mathbf{G}_{j}\in \mathbb{C}^{N\times M}$ denote the channel between the BS and RIS $j$.
Consider the case that all links including BS-user, RIS-user, and BS-user links are operated at a mmWave band (the proposed GNN structure, however, works with any channel model for which it is trained). These links can then be expressed by the following mmWave channel model \cite{mmwave_ch}:
\vspace{-1mm}
\begin{align}\label{ch}
    \mathbf{H}=\frac{1}{\sqrt{L}}\sum_{l=1}^L\alpha_l\mathbf{a}_r\left(\phi_{l}^\mathrm{Rx},\psi_{l}^\mathrm{Rx}\right)\mathbf{a}_t^\ast\left(\phi_{l}^\mathrm{Tx},\psi_{l}^\mathrm{Tx}\right),
\end{align}
where $L$ is the number of paths, and $\alpha_l$ is the complex gain of the $l$-th path. Here, $\phi_{l}^\mathrm{Tx}$, $\psi_{l}^\mathrm{Tx}$, $\phi_{l}^\mathrm{Rx}$, and $\psi_{l}^\mathrm{Rx}$ are the azimuth angle of arrival (AoA), elevation AoA, azimuth angle of departure (AoD), and elevation AoD, respectively.
$\mathbf{a}(\cdot)$ represents the antenna response vector, and the BS and each RIS employs a uniform planar array (UPA) structure to perform 3D beamforming.

The reflection coefficient matrix of RIS $j$ is denoted as $\boldsymbol{\Theta}_j=\mathrm{diag}\{e^{j\theta_{j,1}},\cdots,e^{j\theta_{j,M}}\}$ where $\theta_{j,m}$ is the phase shift of the $m$-th element at RIS $j$.
The composite or effective channel between the BS and user $k$, taking in the reflection off all RISs, can be expressed as
\vspace{-1mm}
\begin{align}\label{eq:comp_ch}
\Tilde{\mathbf{h}}_k^\ast=\mathbf{h}_{k}^\ast+\sum_{j=1}^J\mathbf{f}_{j,k}^\ast\boldsymbol{\Theta}_j\mathbf{G}_{j}=\mathbf{h}_{k}^\ast+\sum_{j\in \mathcal{J}}\mathbf{v}_j^T\text{diag}(\mathbf{f}_{j,k}^\ast)\mathbf{G}_{j},
\end{align}
where $\mathbf{v}_j=[e^{\theta_{j,1}},e^{\theta_{j,2}},\cdots,e^{\theta_{j,M}}]^T$.
Consider downlink communications, the received signal at user $k$ with reflection from all RISs can be written as
\vspace{-1mm}
\begin{align}\label{eq:rx_signal}
y_{k}=&\Tilde{\mathbf{h}}_k^\ast\mathbf{w}_{k} x_k+\sum_{k'=1,  k'\neq k}^K\Tilde{\mathbf{h}}_k^\ast\mathbf{w}_{k'} x_{k'}+n_k,
\end{align}
where $\mathbf{w}_{b,k}\in\mathbb{C}^{N\times 1}$ is the BS transmit beamforming vector for user $k$, $x_k$ is the transmitted symbol for user $k$ with $E[|{x}_k|^2]=1$, and $n_k$ is the additive white Gaussian noise (AWGN) with zero mean and variance $\sigma^2$.

Based on the received signal \eqref{eq:rx_signal}, the signal-to-interference plus noise ratio (SINR) for user $k$ can be written as
\vspace{-1mm}
\begin{align}\label{eq:sinr}
    \gamma_{k}=\frac{\left\|\Tilde{\mathbf{h}}_k^\ast\mathbf{w}_{k} \right\|^2}{\sum\limits_{k'=1, k'\neq k}^K\left\|\Tilde{\mathbf{h}}_k^\ast\mathbf{w}_{k'}\right\|^2 +\sigma^2}.
\end{align}
Since every RIS reflects the signal sent from the BS, the effective channel gain for user $k$ in \eqref{eq:comp_ch} contains the reflected signals from all RISs.
Hence, the RIS phase shifts should be carefully designed to maximize the system performance as each phase shift affects all users in the cell.
In addition, the design of BS beamforming vector is directly dependent on the effective channel.
Therefore, the beamforming vectors for all users and all RIS phase shifts should be jointly designed to maximize the network performance.

\vspace{-1mm}

\subsection{Problem Formulation}
Our goal is to optimize BS beamforming and RIS phase shifts in a multi-RIS assisted wireless network to maximize the network sum rate. This optimization problem can be formulated as
\begin{subequations}\label{P1}
    \begin{align}
        \underset{\mathbf{w}_k, \mathbf{v}_j}{\max} &\mathbb{E} \left[\sum_{k=1}^K\log_2(1+\gamma_k)\right]\\\label{P1b}
    \mathrm{s.t}~&\sum_{k=1}^K\|\mathbf{w}_k\|^2\leq P_t,~|v_{j,m}|=1
    \end{align}
\end{subequations}
where $P_t$ is the total BS transmit power, $v_{j,m}=e^{j\theta_{j,m}}$, and $\gamma_k$ is given in \eqref{eq:sinr}.

Problem \eqref{P1} is challenging to obtain an optimal solution due to its highly non-convex nature.
If we wished to employ a neural network to solve \eqref{P1} in a supervised manner, the labeled data cannot be obtained or generated since  we do not know the optimal solution of \eqref{P1}.
To tackle this problem, we propose a graph based neural network which is trained in an unsupervised manner without labeled data.

\section{Proposed Graph Neural Network Model}

Conventional deep neural network (DNN) is a mapping function from input data to the desirable optimization variables.
However, for problem \eqref{P1}, the input size is directly related to the number of users and RIS nodes, thus the DNN approach yields poor scalability to different network sizes and poor generalization to different network parameter settings.
To improve these scalability and generalization aspects, we propose a GNN architecture to solve \eqref{P1}.
We first construct a graph based on our system model and design a GNN structure to efficiently optimize both BS beamforming and RIS phase shifts.
Channel information will be utilized as network features in a message passing mechanism for sharing information between nodes in the graph.

\subsection{Graph Representation}

\begin{figure}
    \centering
    \includegraphics[width=60mm]{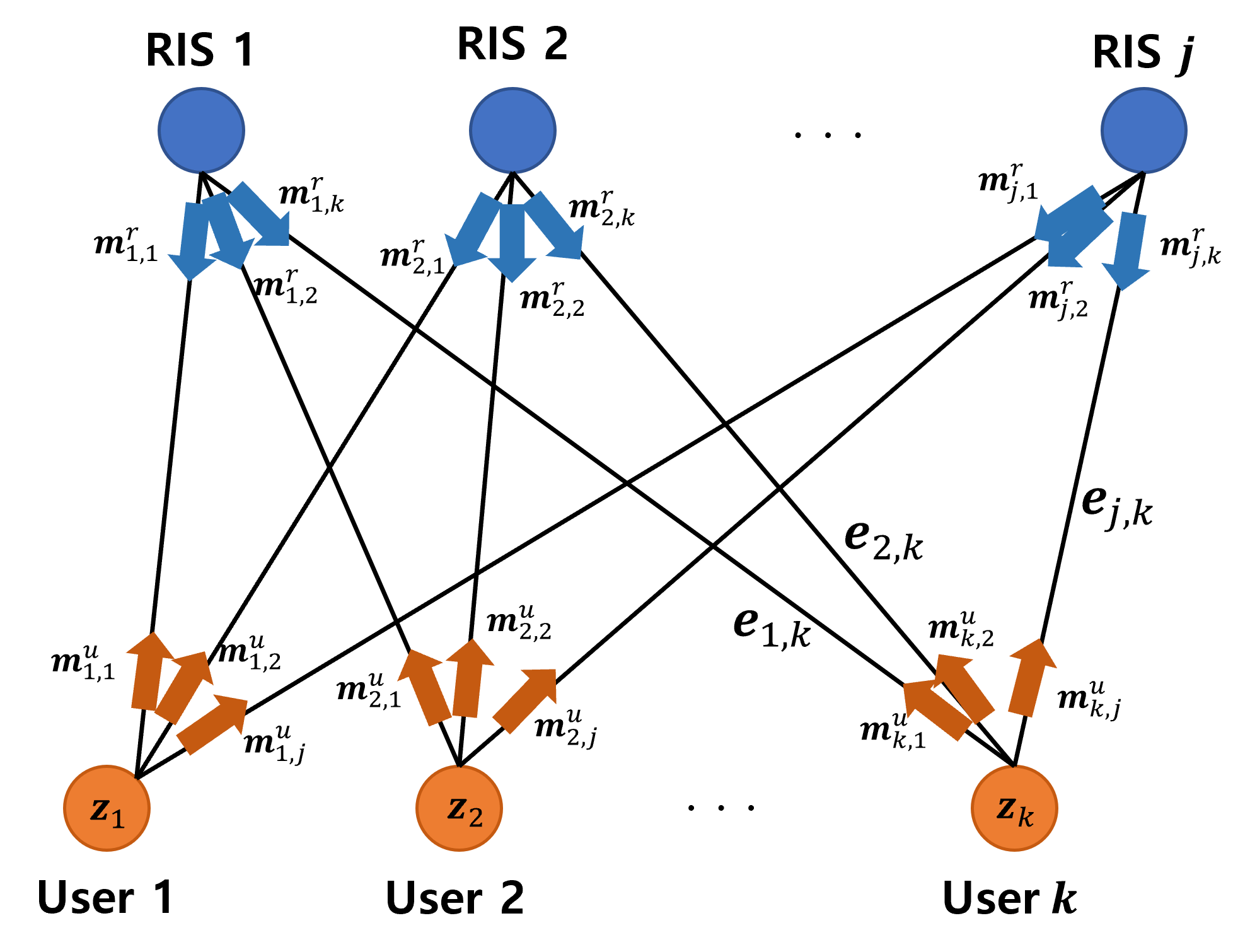}
    \caption{Graph representation of a multi-RIS assisted multi-user network with two types of nodes, where each user node has a feature $\mathbf{z}_k$, and each edge has a feature $\mathbf{e}_{j,k}$. The GNN based on this graph uses message passing mechanism with messages $\mathbf{m}^u_{k,j}$ and $\mathbf{m}^r_{j,k}$.}
    \label{fig:sys}
    \vspace{-2mm}
\end{figure}

We construct an undirected graph $\mathcal{G}=(\mathcal{V},\mathcal{E})$ where $\mathcal{V}$ and $\mathcal{E}$ are the set of nodes and edges in the graph.
  Here, we treat each user as a node of one type and each RIS as a node of another type, where the reflected link from RIS $j$ to user $k$ becomes the edge connecting RIS node $j$ to user node $k$.
Since each RIS reflects a signal from the BS to all users, 
the graph is a bipartite graph, where each RIS node is connected to all user nodes and each user node is connected to all RIS nodes as in Fig. \ref{fig:sys}. (Note there are no edges connecting two users or two RISs directly.)

In the graph, user node $k$ has a feature $\mathbf{z}_k$ as the direct BS channel information, while an RIS node has no feature.
Since each edge in the graph represents the reflected link from an RIS to a user, a feature of the edge $\mathbf{e}_{j,k}$ is chosen as the cascaded channel between RIS node $j$ and user node $k$.
\textcolor{black}{Since wireless channel coefficients are complex, the real and imaginary parts are separated to create the real-valued input vector to a neural network.}
Using the system and channel models in Section \ref{sec:sys}, 
these features can be expressed as
\vspace{-2mm}
\begin{align}\label{eq:feature}\notag
    \mathbf{e}_{j,k}&=\left[\text{vec}\left(\text{Re}\left\{\text{diag}(\mathbf{f}_{j,k}^*)\mathbf{G}_j\right\}\right)^T, \text{vec}\left(\text{Im}\left\{\text{diag}(\mathbf{f}_{j,k}^*)\mathbf{G}_j\right\}\right)^T \right]^T,\\\notag    \mathbf{z}_k&=\left[\text{Re}\left\{\mathbf{h}_k\right\}^T, \text{Im}\left\{\mathbf{h}_k\right\}^T\right]^T.
\end{align}

The optimization variable of each RIS node is the RIS phase shifts, and each user node is the beamforming vector from the BS.
Denote the representation vector associated with user node $k$ as $\mathbf{d}_k$ and with RIS node $j$ as $\mathbf{c}_j$. These vectors are updated by the GNN structure and subsequently mapped into the desired optimization variables.
Our goal is to update every representation vector in the graph and train the GNN in such a way that the RIS phase shift $\mathbf{v}_j$ can be obtained from $\mathbf{c}_j$ and the beamforming vector $\mathbf{w}_k$ can be obtained from $\mathbf{d}_k$.

\vspace{-3mm}
\subsection{GNN Architecture}
If we only exploit the node and edge features to optimize the BS beamforming and RIS phase shifts, the interference introduced from other nodes cannot be considered because these features as in \eqref{eq:feature} only include information about a direct link but not an interference link.
To capture the intra-cell interference between users and the additional 
interference caused by the reflected signals from RISs, the interaction between nodes is embedded in the GNN structure to make the GNN learn to deduce the interference.
Specifically, we employ a message passing mechanism described below to capture nodes interaction and update the representation vectors of user nodes and RIS nodes.

The operation of a GNN architecture with message passing is composed of 2 steps: 1) message generation, and 2) representation vector update.
Since there are two types of nodes in the graph, user nodes and RIS nodes, we need to consider two types of messages: the message from an RIS to a user, and the message from a user to an RIS.

\textit{1) Message generation:}
At the $t$-th layer in the GNN, the messages from RIS $j$ to user $k$ and from user $k$ to RIS $j$, respectively, can be expressed as
\begin{align}\notag
\mathbf{m}_{k,j}^{u,(t)}&=f_1^u\left(\mathbf{z}_k,\mathbf{e}_{j,k},\mathbf{d}_k^{(t-1)},\phi\left(\{\mathbf{m}_{j,k}^{r,(t-1)}\}_{j\in \mathcal{J}}\right)\right),\\
\mathbf{m}_{j,k}^{r,(t)}&=f_1^{r}\left(\mathbf{e}_{j,k},\mathbf{c}_j^{(t-1)},\phi\left(\{\mathbf{m}_{k,j}^{u, (t)}\}_{k\in \mathcal{K}}\right)\right),
\end{align}
where each function $f_1^u(\cdot)$ and $f_1^r(\cdot)$ is a fully connected neural network for the user node and the RIS node, respectively, and  $\phi(\cdot)$ is the element-wise mean function to aggregate messages from the neighboring (connected) nodes.
Note that the function $\phi(\cdot)$ needs to act only on each node's received messages, which are collected from all the nodes of other type connected to this node. 
After constructing the messages at each node, the node sends each message to the respective neighboring node.

\textit{2) Representation vector update:}
After receiving the messages, each node aggregates the messages from neighboring nodes and updates its representation vector as follows.
\begin{align}\notag
    \mathbf{d}_k^{(t)}&=f_2^u\left(\mathbf{z}_k,\mathbf{d}_k^{(t-1)},\phi\left(\{\mathbf{m}_{j,k}^{r,(t)}\}_{j\in\mathcal{J}}\right)\right),\\
    \mathbf{c}_j^{(t)}&=f_2^r\left(\mathbf{c}_j^{(t-1)}, \phi\left(\{\mathbf{m}_{k,j}^{u,(t)}\}_{k\in \mathcal{K}}\right)\right),
\end{align}
where $f_2^u(\cdot)$ and $f_2^r(\cdot)$ are fully connected neural networks at the user node and the RIS node, respectively.
The messages arriving at each node contain information from other nodes, including the beamforming vector for other users and the RIS phase shifts of all RISs, which facilitate the inference of the aggregated interference at user $k$.
Therefore, by exchanging messages among neighboring (connected) nodes, each node can update its representation vector with the implicit knowledge of unconnected nodes.

After updating the representation vectors for $T$ layers in the GNN, the resulting representation vectors $\mathbf{d}_k^{(T)}$ and $\mathbf{c}_j^{(T)}$ are projected into the feasible region of beamforming constraint and the unit modulus constraint of RIS phase shift as in \eqref{P1b}, respectively.
\textcolor{black}{The complex-valued BS beamforming and RIS phase shifts vectors can be obtained from the real-valued representation vectors as follows.}
\begin{align}\label{eq:projecting}\notag
\mathbf{w}_k&=\sqrt{\frac{Pt}{K}}\frac{\mathbf{d}_k^{(T)}(1:N)+j\mathbf{d}_k^{(T)}(N+1:2N)}{\|\mathbf{d}_k^{(T)}(1:N)+j\mathbf{d}_k^{(T)}(N+1:2N)\|_2},\\
v_{j,m}&=e^{jc_{j,m}^{(T)}}.
\end{align}
Note that the representation $c_{j,m}^{(T)}$ is not restricted to be within the range of $[-\pi, \pi]$, but the projection of $c_{j,m}^{T}$ onto $v_{j,m}$ results in a feasible phase shift solution for \eqref{P1}.

\subsection{Training and Implementation}

In the proposed GNN structure, the neural network weights $\Omega_1^u$ in $f_1^u(\cdot)$, $\Omega_1^r$ in $f_1^r(\cdot)$, $\Omega_2^u$ in $f_2^u(\cdot)$, and $\Omega_2^r$ in $f_2^r(\cdot)$ need to be trained to maximize the sum rate in an unsupervised manner.
In the off-line training phase, we train the GNN parameters for message generation and representation vector update to minimize the following loss function:
\begin{align}\label{eq:loss}
    L(\boldsymbol{\Omega})=-\mathbb{E}_{\mathcal{B}} \left[\sum_{k=1}^K\log_2(1+\gamma_k)\right],
\end{align}
where $\mathcal{B}$ is the mini-batch set and $\boldsymbol{\Omega}=\{\boldsymbol{\Omega}_1^u,\boldsymbol{\Omega}_1^r,\boldsymbol{\Omega}_2^u,\boldsymbol{\Omega}_2^r\}$ is the sef of parameters.
For each training sample in a mini-batch, the channels of users and RISs are generated randomly based on the model in Section II and are used to create the GNN features in (6). The SINR $\gamma_k$ as in \eqref{eq:sinr} is calculated from the generated channels and the GNN outputs as in (9).

The minimization of empirical loss in \eqref{eq:loss} can be achieved by a mini-batch stochastic geometry descent (SGD) method which updates the parameters at iteration $i$ as
\begin{align}
    \boldsymbol{\Omega}^{(i+1)}\leftarrow \boldsymbol{\Omega}^{(i)}-\eta \nabla_{\boldsymbol{\Omega}}\mathbb{E}_{\mathcal{B}}\left[\sum_{k=1}^K\log_2(1+\gamma_k)\right],
\end{align}
where $\eta$ is the learning rate.

After training these neural networks, each user node and each RIS node has a copy of two trained networks, that is, $f_1^u(\cdot)$ and $f_2^u(\cdot)$ for the user node, and $f_1^r(\cdot)$ and $f_2^r(\cdot)$ for the RIS node. Note that these trained neural networks are the same at all nodes of the same type. As such, the GNN structure provides excellent scalability by sharing the same trained neural network structure at all nodes of the same type.

In the on-line implementation phase, current channel state information including the direct and cascaded channels of all users are first estimated and are used to create the GNN features as in \eqref{eq:feature}.
Then, each RIS node $j$ generates the messages $\mathbf{m}_{j,k}$ for all $k$ using $f_1^r(\cdot)$ and passes each message to the corresponding user node $k$.
Each user node $k$ also generates the messages for all RIS nodes using $f_1^u(\cdot)$ after aggregating the messages received from neighboring RIS nodes, then passes the corresponding messages to the connected RIS nodes.
The representation vector at each user node and each RIS node is then updated using the trained neural network $f_2^u(\cdot)$ and $f_2^r(\cdot)$.
This procedure is iteratively performed for $T$ layers and the final representation vectors are projected into the feasible output region as in \eqref{eq:projecting}.


\section{Simulation Results}\label{simul}
In this section, we evaluate the proposed GNN structure for a multi-RIS assisted single-cell wireless network.
We use the carrier frequency and bandwidth of 28GHz and 1GHz, respectively, and the noise power spectral density of -174dBm/Hz.
In this network, 4 user nodes and 6 RIS nodes are uniformly located in a region of $200 \times 200 \textrm{m}^2$.
The BS employs an $4\times4$ UPA antenna, and the total transmit power is set to 20dBm.
We adopt the mmWave path loss model measured in New York City and chose the path-loss parameters as in\cite{NY_prob}. 
The neural networks at each node have 2 hidden layers with 512 hidden nodes and are trained via the Adam optimizer with a learning rate of $10^{-4}$.
In each training epoch, 100 samples are used to construct a mini-batch, where the channel and location of users and RISs are randomly generated for each sample, and 100 iterations are run for each mini-batch to update the parameters of each neural network in an epoch.
In the simulations, we assume perfect channel information for on-line GNN implementation and use $T=2$ layers.

\begin{figure}
    \centering
    \includegraphics[width=60mm]{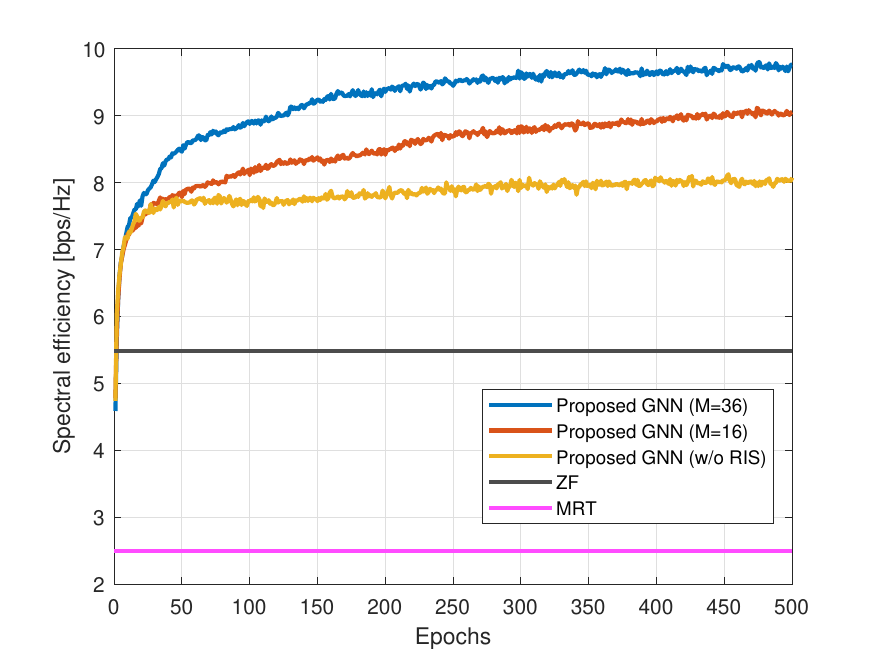}
    \caption{Convergence behavior of the proposed GNN training.}
    \label{fig:conv}
    \vspace{-3mm}
\end{figure}

Fig. \ref{fig:conv} shows the convergence behavior of the proposed GNN structure.
In the wireless network without RIS, only the BS beamforming vectors for users are optimized, thus the network performance is worse than a multi-RIS assisted network.
Note also that while a larger number of passive elements at each RIS leads to higher network spectral efficiency as shown in Fig. \ref{fig:conv}, it requires a longer training time for convergence since more variables need to be learned from the network.

\begin{figure}
    \centering
    \includegraphics[width=60mm]{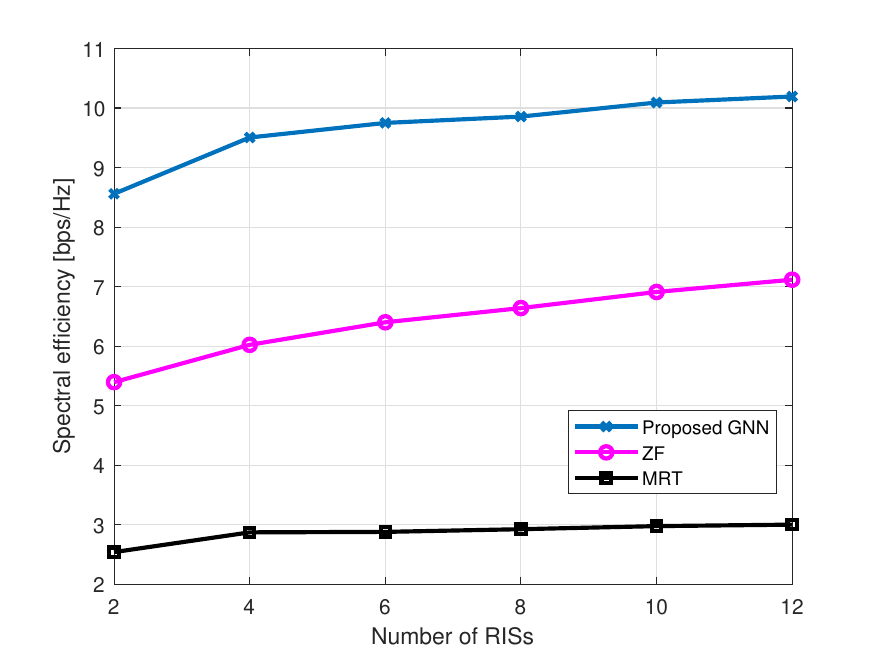}
    \caption{Generalization performance of the proposed GNN trained with $K=4$ and $J=6$.}
    \label{fig:general}
    \vspace{-5mm}
\end{figure}

Fig. \ref{fig:general} demonstrates the scalability of the proposed GNN architecture.
For benchmark, we adopt (1) MRT beamforming while aligning each RIS's phase shifts to a user having the maximum effective channel gain \cite{Limwcnc}, and (2) ZF beamforming with the RIS phase shift optimization in \cite{WCL_RIS_cellfree}.
Using the trained neural networks in Fig. \ref{fig:conv}, performance of the GNN is evaluated while changing the number of RISs.
As the number of RISs increases, we can observe that the GNN always outperforms the conventional benchmark algorithms.
Note that a DNN structure will not have this scalability without re-designing and retraining the DNN for more RISs.
Furthermore, since we employ the message passing mechanism, interference information is exchanged among nodes, making it possible to obtain the optimal solution via the proposed GNN architecture.
This result demonstrates its excellent scalability and superior performance obtained by the message passing mechanism.

\vspace{-2mm}
\section{Conclusion}\label{conclu}
We proposed a novel graph neural network (GNN) architecture to joinlty optimize BS beamforming and RIS phase shifts in a multi-RIS assisted single-cell wireless network.
Specifically, we employ a message passing mechanism to facilitate the interaction among nodes, which plays a key role in optimizing the network sum rate performance.
Simulation results demonstrate that the proposed GNN structure provides strong scalability and good generalization to different network sizes and parameters, while achieving superior performance compared to conventional schemes.

\bibliographystyle{IEEEtran}

\end{document}